\documentclass[aps,showpacs,pra,twocolumn,superscriptaddress]{revtex4}
\usepackage{epsfig,amsmath}
\usepackage{subfigure}
\usepackage{graphicx}
\usepackage{dcolumn}
\usepackage{stmaryrd}
\usepackage{mathrsfs}
\usepackage{pifont}
\usepackage{amsthm}
\usepackage{amssymb}
\usepackage{bm}
\usepackage{latexsym}
\usepackage{hyperref}

\newcommand{\beq}{\begin{equation}}
\newcommand{\eeq}{\end{equation}}
\newcommand{\beqa}{\begin{eqnarray}}
\newcommand{\eeqa}{\end{eqnarray}}

\begin{document}

\title{An adiabatic Leakage Elimination Operator in experimental framework}
\author{Zhao-Ming Wang}
\affiliation{Department of Physics, Ocean University of China, Qingdao 266100, China}
\affiliation{Department of Theoretical Physics and History of Science, University of the
Basque Country UPV/EHU, 48008, Spain}
\author{Mark S. Byrd}
\affiliation{Department of Physics, Southern Illinois University, Carbondale,
Illinois 62901-4401, USA}
\author{Jun Jing}
\affiliation{Institute of Atom and Molecular Physics and Jilin Provincial Key Laboratory
of Applied Atomic and Molecular Spectroscopy, Jilin University, Changchun
130012, Jilin, China}
\author{Lian-Ao Wu \footnote{%
Corresponding author: lianao.wu@ehu.es}}
\affiliation{Department of Theoretical Physics and History of Science, University of the
Basque Country UPV/EHU, 48008, Spain}
\affiliation{IKERBASQUE, Basque Foundation for Science, 48011 Bilbao, Spain}
\date{\today }

\begin{abstract}
Adiabatic evolution is used in a variety of quantum information processing 
tasks.  However, the elimination of errors is not as well-developed as it is for circuit model processing.  
Here, we present a strategy to accelerate a reliable quantum adiabatic process by adding
Leakage Elimination Operators (LEO) to the evolution which are a sequence of
pulse controls acting in an adiabatic subspace. Using the Feshbach $PQ$
partitioning technique, we obtain an analytical solution which traces the
footprint of the target eigenstate. The effectiveness of the
LEO is independent of the specific form of the pulse but depends on the
average frequency of the control function. Furthermore, we give the exact
expression of the control function in an experimental framework by a counter
unitary transformation, thus the physical meaning of the LEO is clear.
Our results reveal the equivalence of the control function between two
different formalisms which aids in implementation.
\end{abstract}

\maketitle


\section{Introduction}

Adiabatic theorem states that a system prepared in a nondegenerate
eigenstate will remain in that instantaneous eigenstate if the evolution time
is infinitely long even thought the eigenvalue itself could change.  
The performance of the adiabatic evolution, including limiting the rate of 
change of the eigenvalue is
dictated by a long evolution time compared to the inverse of a power of the
energy gap \cite{Messiah,Jansen,Sarandy}. It plays an important role in the
quantum information processing, such as quantum state transmission \cite%
{Srinivasa2007,Balachandran2008,Farooq}, adiabatic quantum computation \cite%
{Childs,Sarandy2005,Zanardi,Pyshkin}, adiabatic quantum algorithms \cite%
{Farhi,Garnerone,WangHF},  and heat transfer \cite{Li10,Wu09}. During the system evolution, the dissipation
and noise always exist and decoherence and leakage from one eigenstate to 
another accumulated during a long
runtime may destroy the accuracy of the system. Therefore, several
schemes have been proposed to speed up adiabatic passage \cite%
{Bergmann,Kral,Muga} while reducing errors.  One method is 
to modify 
the original Hamiltonian to compensate for nonadiabatic errors, which is the
so-called transitionless driving \cite{Berry}, counteradiabatic control \cite%
{Demirplak}, or higher order invariants \cite{Demirplak2008}. Another method is 
to apply a sequence of fast pulses during the dynamical process, 
consequently the adiabaticity \cite{JingPRA2014}, the adiabatic quantum
computation \cite{WangHF,Pyshkin}, and non-adiabatic quantum state
transmission \cite{Oh01,Wang2016} can be sped up.  

Leakage Elimination Operators (LEOs) are a type of dynamical decoupling 
control \cite{Viola0} that were introduced to 
specifically counteract leakage in a two-state system which 
encodes one logical qubit in a multilevel Hilbert space \cite%
{WuPRL2002,Byrd,Campo,Campo2011}. In general, the total Hamiltonian can be
written as $H=H_{P}+H_{Q}+H_{L}$, where $H_{P}$ acts on the 
subspace of interest (e.g. the logical subspace), $P$, $H_{Q}$ acts on the remaining
Hilbert subspace orthogonal to the $P$ subspace, $Q$ and $H_{L}$ is the part of the Hamiltonian that can cause 
transitions between $P$ and $Q$. If an operator satisfies $\{R_{L},H_L\}=0$, and 
$[R_{L},P]=[R_{L},Q]=0$, then $R_{L}$ is an effective LEO for the system. It satisfies 
the $\lim_{m\rightarrow \infty }[e^{-iHt/m}R_{L}^{\dag
}e^{-iHt/m}R_{L}]^{m}=e^{-iH_{P}t/m}e^{-iH_{Q}t/m}$ \cite{JingPRL2015}. This 
can eliminate the transition from $P$ to $Q$.  Often unbounded
fast and strong pulses (Bang-Bang) are sought to eliminate errors 
\cite{Viola,Vitali}.  However, such pulses are an idealization 
which is unattainable in experiment.  Furthermore, such Bang-Bang sequences 
has been shown to be unnecessary; the effectiveness of LEOs depends only on the 
exponential of the integral of the pulse sequence in the time domain \cite{JingPRL2015, Pyshkin}.  

During the adiabatic evolution, the transition from one instantaneous
eigenstate to another always ruins the adiabaticity. Can we add an LEO in
one of the instantaneous eigenstate subspaces to prevent this transition? In
this paper, we propose such a scheme to speed up the adiabatic quantum
evolution by introducing an LEO in the adiabatic framework of the
Hamiltonian. For two simple examples, by using the Feshbach $PQ$
partitioning technique \cite{Wu2009} we find that the transitions are
greatly suppressed via an external LEO control even in a non-adiabatic
regime. By using an appropriate unitary transformation, we provide a description of
the control function in an experimental framework. Furthermore, the
calculation shows that the fidelity is only determined by the
average frequency of the control function, this greatly expands the choice of
the types of pulses.

\section{LEO in an adiabatic framework}

Given a time dependent Hamiltonian $H(t)$, the instantaneous eigenstate $|
E_{n}(t)\rangle $ and the corresponding eigenvalue are given by,
\begin{equation}
H(t)| E_{n}(t)\rangle =E_{n}(t)| E_{n}(t)\rangle .  \label{eq:01}
\end{equation}
At any particular instant the constitute a complete orthonormal set $\langle
E_{n}(t)| E_{m}(t)\rangle =\delta _{nm}$.  They provide a general
solution to the time-dependent Schr\"{o}dinger equation $i\overset{\cdot }{|
\Psi \rangle }=H| \Psi\rangle $ as a wave function
$| \Psi\rangle $ that can be expressed as a linear combination in the 
adiabatic time-dependent basis
\begin{equation}
|\Psi \rangle =\sum\limits_{n}c_{n}(t)| E_{n}(t)\rangle,
\end{equation}
or time-independent basis
\begin{equation}
| \Psi \rangle =\sum\limits_{n}d_{n}(t)| E_{n}(0)\rangle.
\end{equation}

A unitary transformation can be used to transform from the time-independent basis
to the adiabatic basis,
\begin{equation}
U(t)=\sum\limits_{k}| E_{k}(t)\rangle \langle E_{k}|.
\end{equation}
So at any particular instant, $U(t)$ maps the time-independent state 
$|E_{k}\rangle $ onto the time-dependent state $|E_{k}(t)\rangle $. The
corresponding gauge transformation of the Hamiltonian is
\begin{eqnarray}
H_{a}(t) &=&U^{^{\dag }}H_{e}(t)U-iU^{^{\dag }}\overset{\cdot }{U}  \notag \\
&=&H_{d}(t)+M(t),
\end{eqnarray}%
where $H_{e}$ and $H_{a}$ are the representation of the Hamiltonian in an
experimental (lab) frame and adiabatic frame, respectively. The diagonal
terms are 
\begin{eqnarray*}
H_{d}(t) &=&U^{^{\dag }}H_{e}(t)U \\
&=&diag(E_{0}(t)-i\langle E_{0}(t)|\overset{\cdot }{E_{0}}(t)\rangle , \\
&&E_{1}(t)-i\langle E_{1}(t)|\overset{\cdot }{E_{1}}(t)\rangle ,\ldots ),
\end{eqnarray*}%
and the off-diagonal terms, responsible for transformations, are 
\begin{eqnarray*}
M_{mn}(t) &=&-iU^{^{\dag }}\overset{\cdot }{U} \\
&=&-i\langle E_{m}(t)|\overset{\cdot }{E_{n}}(t)\rangle (m\neq n).
\end{eqnarray*}%

In what follows we will consider a transitionless process
during the dynamics in a non-adiabatic regime. Our strategy is to add an LEO control into the original Hamiltonian
\begin{equation}
H_{LEO}=f(t)|E_{0}(t)\rangle \langle E_{0}(t)|,
\end{equation}%
where $f(t)$ is the control function which describes a sequence of fast
pulses.  This will be describe an LEO can be used to reduce errors from an encoded
(logical) subspace to the rest of the system subspace whether the pulses are the
ideal pulses (bang-bang controls) \cite{Wu2009} or non-ideal
pulses \cite{JingPRL2015}.  In contrast to adding LEOs directly into an
lab frame \cite{JingPRL2015,Wu2009}, here we add
an LEO in an adiabatic frame. The transition from one eigenstate to
other subspaces is prevented during the evolution. Then if the control
function is fast and strong enough, the system evolution will behave as though
it is adiabatic even in a non-adiabatic regime.  

We now illustrate the adiabatic LEO contorl by a simple two level system (Example 1).  
The Hamiltonian reads
\begin{equation}
H_{0}(t)=\frac{\omega _{1}}{2}[\cos (\omega t)\sigma _{z}+\sin (\omega
t)\sigma _{x}],  \label{Eq0}
\end{equation}%
where $\cos(\omega t)$ and $\sin(\omega t)$\ describe a field whose direction
changes from $z$ to $x$ at constant angular velocity $\omega $.  Its
instantaneous eigenvalues are $E_{0}=-\omega _{1}/2,E_{1}=\omega _{1}/2$ and
the eigenvectors of $H_0$ can be expressed as
\begin{eqnarray}
\left\vert E_{0}(t)\right\rangle &=&-\sin \omega t/2|\uparrow \rangle +\cos
\omega t/2|\downarrow \rangle , \\
\left\vert E_{1}(t)\right\rangle &=&\cos \omega t/2|\uparrow \rangle +\sin
\omega t/2|\downarrow \rangle.
\end{eqnarray}

The Hamiltonian in Eq.~(\ref{Eq0}) in an adiabatic basis, with the LEO control, can be
written as
\begin{equation}
H(t)=\left(
\begin{array}{cc}
-\omega _{1}/2+f(t) & \frac{-i\omega }{2} \\
\frac{i\omega }{2} & \omega _{1}/2%
\end{array}%
\right) .  \label{Eq1}
\end{equation}
Without loss of generality, in Eq.~(\ref{Eq1}), we let $H_{0}^{^{\prime
}}=H_{0}-\omega _{1}/2$, i.e., we change the energy zero point energy. 
Ideally, if we
turn on a strong, fast control $f(t)\propto \delta (t-n\tau )$ at 
times $n\tau $ ($n=0,1,\ldots $), the LEO control generates to the LEO $R_{L}$ \cite%
{WuPRL2002} in the adiabatic framework, or an adiabatic LEO. This operator
satisfies $\{R_{L},H_L\}=0,$ and 
\begin{equation}
e^{-iH(n\tau )\tau}R_{L}^{\dag }e^{-iH((n-1)\tau)\tau}R_{L}\approx e^{-iH_{d}(n\tau
+(n-1)\tau )\tau}.
\end{equation}
When $\tau \rightarrow 0$ and $t\approx n\tau $, this Bang-Bang corresponds to 
a parity-kick sequence and eliminates the leakage $H_L$. Furthermore, all leakage such as $LB$ can
be eliminated by $R_{L}$, where $B$ can be an operator of another system, such
as an external bath \cite{WuPRL2002}. When $f(t)=\delta (t-n\tau )$, 
a parity-kick at $=n\tau $ corresponds to the rotation LEO $R_{L}=-iZ$, 
such that $R_{L}^{\dagger}H_{L}R_{L}=-H_{L}$ and $H_{L}$ is removed.

\section{Results and Discussions}

We have constructed the adiabatic LEO and next we will analyze the potential 
speedup of the evolution. Using the $PQ$ partitioning technique, an 
$n$-dimensional wave
function $\psi$ can be divided into two parts: a one-dimensional vector
of interest $P(t)$ and the rest $(n-1)$-dimensional vector $Q(t)$. $\psi,H$ can be written as
\begin{equation}
\psi=\left[
\begin{array}{c}
P \\
Q%
\end{array}%
\right] ,\quad H_{P}+H_Q=\left[
\begin{array}{cc}
h & 0 \\
0 & D%
\end{array}%
\right] ,\quad H_{L}=\left[
\begin{array}{cc}
0 & R \\
W & 0%
\end{array}%
\right] ,
\end{equation}%
where the $1\times 1$ matrix $h$ and $(n-1)\times (n-1)$ matrix $D$ are the
self-Hamiltonians in the subspaces of $P$ and $Q$. For our example, $h$ $%
=$ $f(t)-\omega _{1}/2$ , $R=\frac{-i\omega }{2}$, $W=\frac{i\omega }{2}$
and $D=\omega _{1}/2$.  In the selected one dimensional subspace, $p$
satisfies
\begin{equation}
\overset{\cdot }{p}=\int\nolimits_{0}^{t}g^{^{\prime }}(t,s)p(s)ds,
\label{Eq9}
\end{equation}%
where we have used $p(t)=\exp [i\int\nolimits_{0}^{t}h(s^{^{\prime
}})ds^{^{\prime }}]P(t)$. For the Hamiltonian in Eq.~(\ref{Eq1}), the propagator
$g^{^{\prime }}(t,s)=-g(t,s)\exp [i\int\nolimits_{s}^{t}h(s^{^{\prime
}})ds^{^{\prime }}]$, $g(t,s)=R(t)G(t,s)W(s)$. $G(t,s)=\Gamma _{\leftarrow
}\{\exp [-i\int\nolimits_{s}^{t}D(s^{^{\prime }}){ds^{^{\prime }}}]\}$ is a
time-ordered evolution operator. Specifically, 
\begin{eqnarray}
g^{\prime }(t,s) &=&\frac{-\omega ^{2}}{4}\exp \{{i\int%
\nolimits_{s}^{t}[f(s^{^{\prime }})-\omega _{1}]ds^{^{\prime }}\}}  \notag \\
&=&\frac{-\omega ^{2}}{4}\exp \{{i[(}\left\langle \omega
_{2}(s,t)\right\rangle -{\omega _{1}})(t-s)]\}{,}  \label{Eq10}
\end{eqnarray}%
where we have defined the average control frequency
\begin{equation}
\left\langle \omega _{2}(s,t)\right\rangle =[{\int\nolimits_{s}^{t}f(s^{^{%
\prime }})ds^{^{\prime }}]/}(t-s),  \label{Eq11}
\end{equation}%
from time $s$ to time $t$. The adiabatic path requires $\overset{\cdot }{p}%
=0 $, i.e., the eigenstate population of the time-dependent Hamiltonian $H(t)
$ is constant in time. Clearly when $\omega \rightarrow 0$, $\overset{\cdot }%
{p}=0$, the standard adiabatic condition is satisfied. For finite $\omega $,
$g^{\prime }(t,s)$ is a quickly oscillating periodic function.  Its frequency
can be enhanced by increasing the average control frequency $\left\langle
\omega _{2}\right\rangle $. Then the integrand in Eq.~(\ref{Eq9}) is a
product of a quickly oscillating function $g^{\prime }(t,s)$ and a slowly
varying function $p(s)$. According to the Riemann-Lebesgue lemma, the
integral of the product of a fast varying and slowly varying function
averages to approximately zero. The integral is more inclined to be zero for a
a larger $\left\langle \omega _{2}\right\rangle $. The effectiveness of the
LEO depends on the average frequency of the control function $f(t)$, but 
does not depend on the details of $f(t)$.

\begin{figure}[t]
\centering
\includegraphics[width=0.8 \columnwidth]{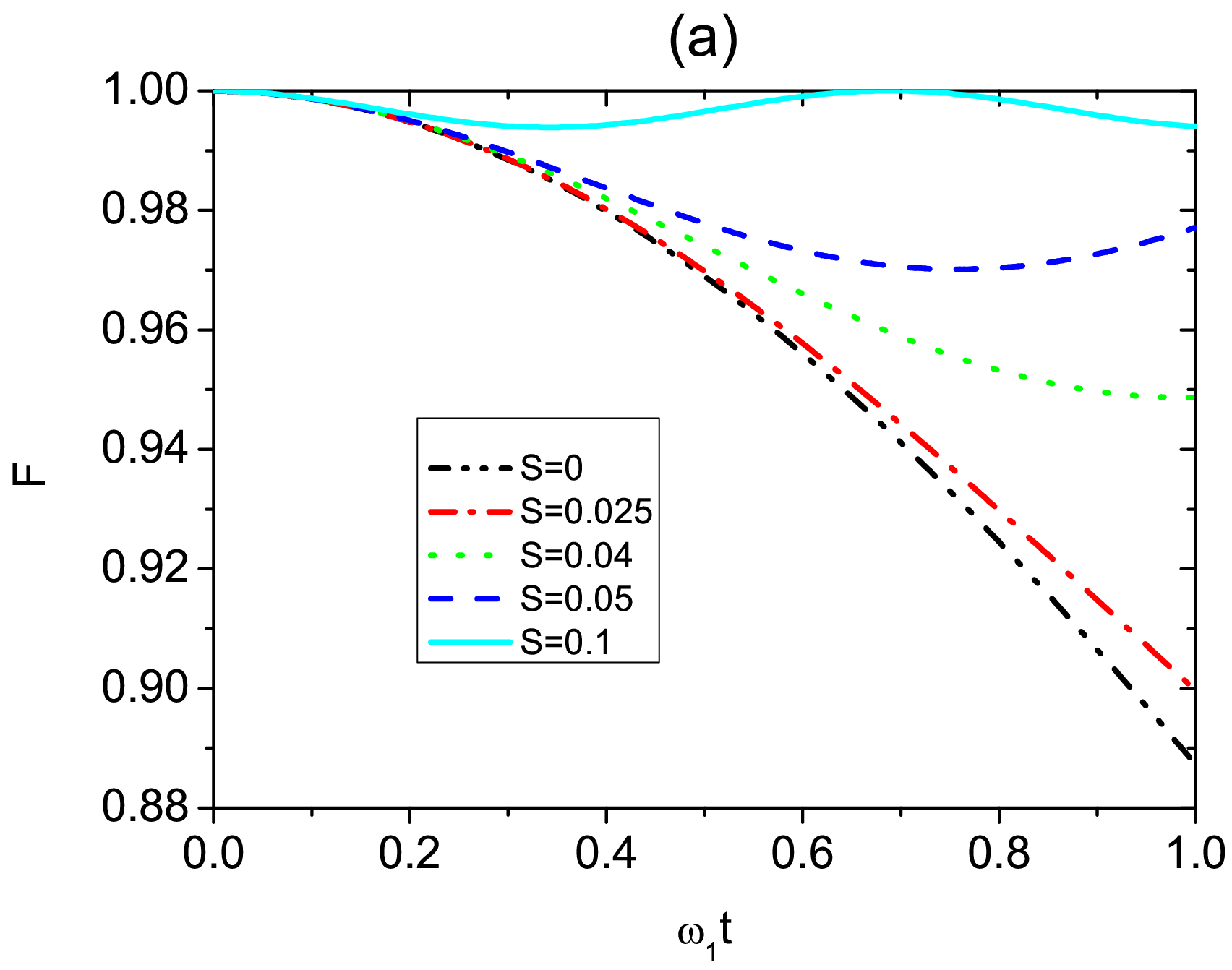} %
\includegraphics[width=0.8 \columnwidth]{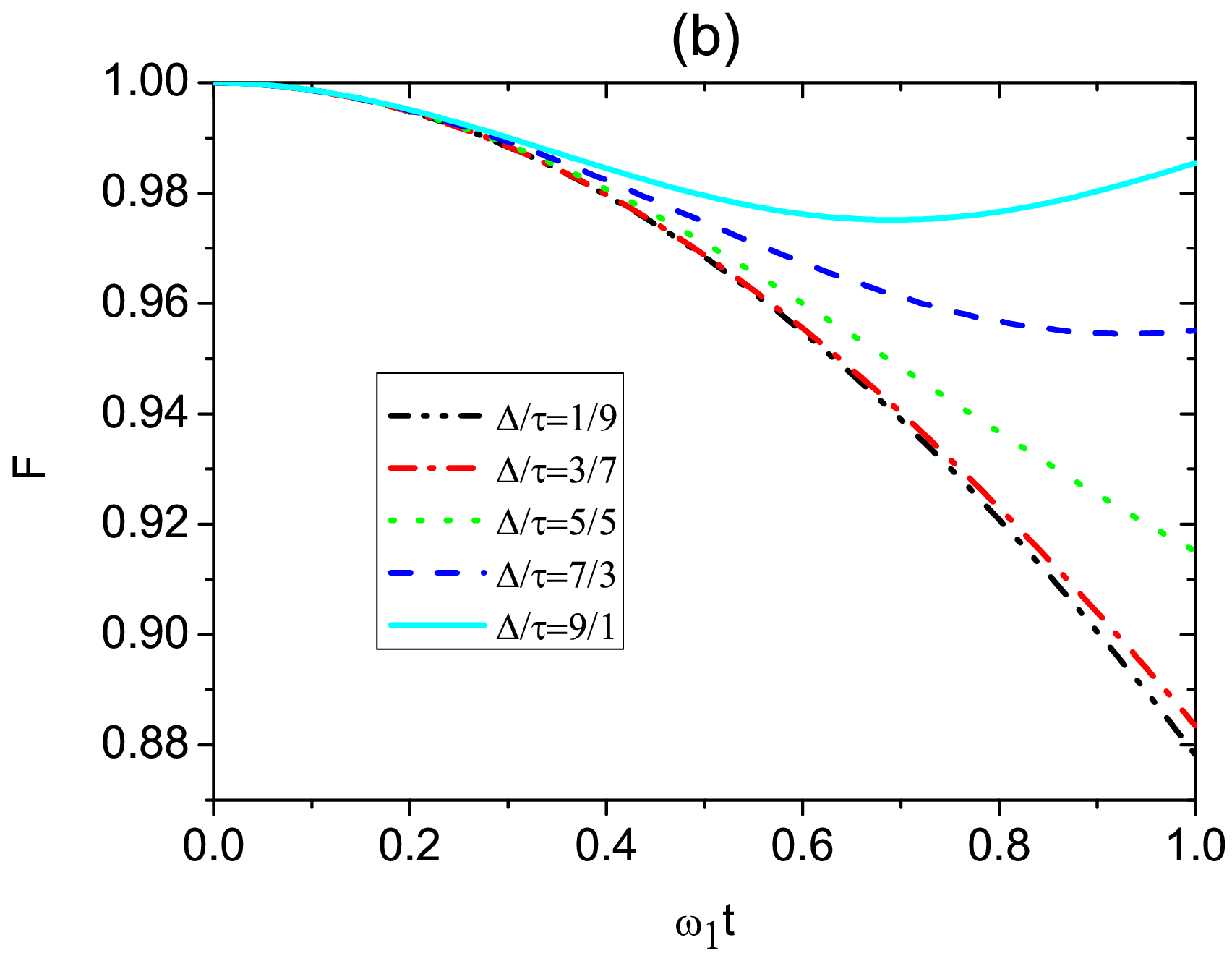}
\caption{(Color on line) Example 1: Fidelity versus parameter $\protect\omega%
_{1}t$ for (a) different integral $S$, $\Delta=\protect\xi,\Delta/\protect%
\tau=1/1,T=1/\protect\omega_{1}$, the time step length we used to calculate
is taken as $\protect\xi=0.005/\protect\omega_{1}$; (b) different ratio of $%
\Delta/\protect\tau$, $\Delta+\protect\tau=10\protect\xi$, $S=0.03$.}
\label{fig:1}
\end{figure}

\begin{figure}[t]
\centering
\includegraphics[width=0.8 \columnwidth]{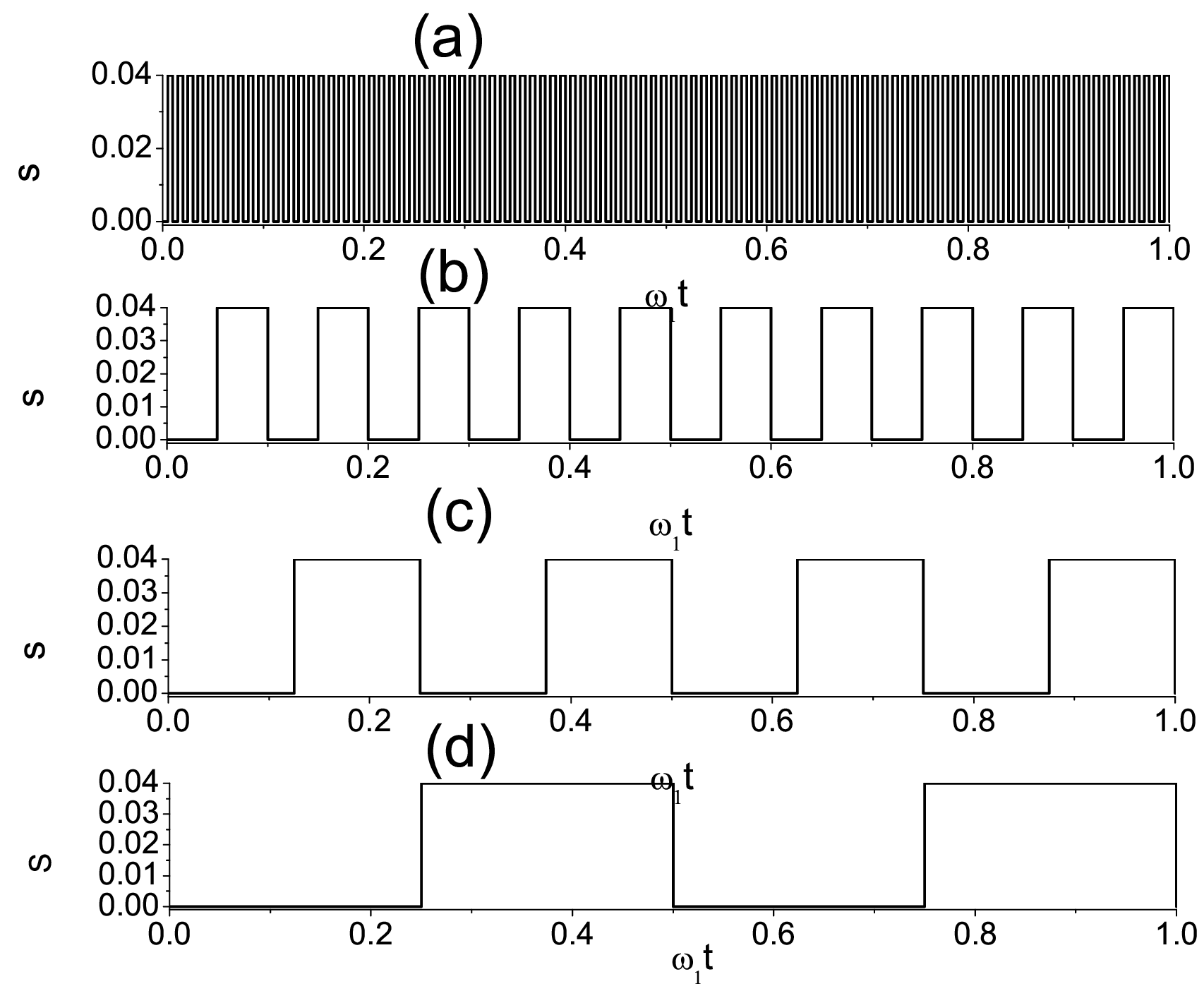} %
\includegraphics[width=0.8 \columnwidth]{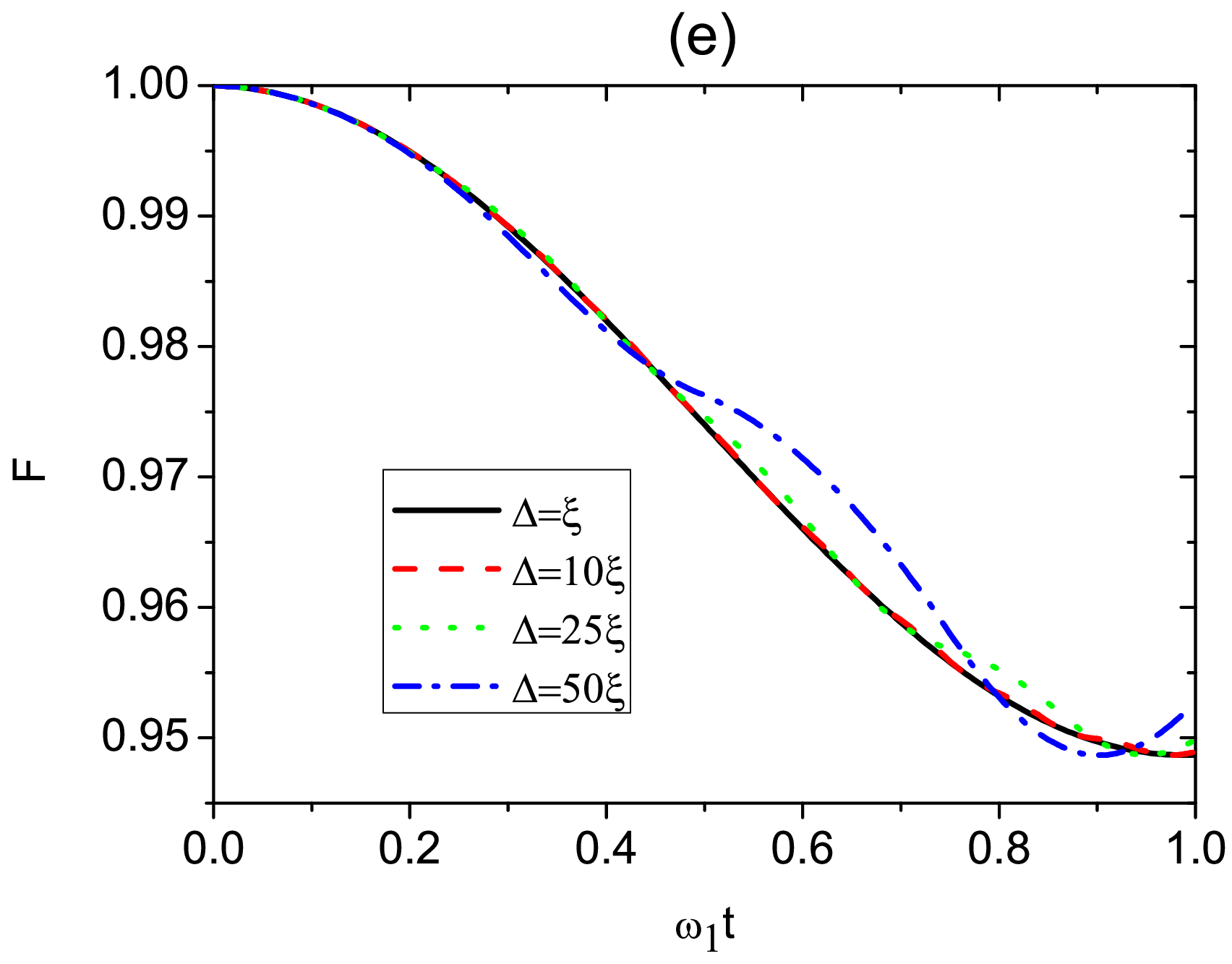}
\caption{(Color on line) Example 1: Fidelity versus parameter $\protect\omega%
_{1}t$ for different pulse interval, (a) $\Delta=\protect\xi$; (b) $\Delta=10%
\protect\xi$; (c)$\Delta=25\protect\xi$; (d) $\Delta=50\protect\xi$. $\Delta/%
\protect\tau=1/1.$}
\label{fig:2}
\end{figure}

Now we present the numerical calculation results. Suppose $f(t)$ is chosen
as a sequence of rectangular pulses, $f(t)=I$ with control and $f(t)=0$
without control. $I$ is the pulse strength. $\tau(\Delta)$ is the time
interval of the free evolution (under control). For a regular rectangular
pulse, $\Delta/\tau$ is a constant.

In our numerically calculation we use the time step length $\xi
=0.005/\omega_{1}$ to calculate the propagator in Example 1. The integral $%
S=\int\nolimits_{n\xi }^{(n+1)\xi }f(s^{^{\prime }})ds^{^{\prime }}$ ($%
n=0,1,2,...$) in the propagator determines the controllability from Eq.~(\ref%
{Eq10}). Note that $S$ is a dimensionless parameter.

Suppose the system is initially in the ground state of $H(0)$. The fidelity
is defined as $F=\left\vert \left\langle \Psi (t)|E_{0}(t)\right\rangle
\right\vert $, where $|E_{0}(t)\rangle $ is the instantaneous ground state
in Eq.~(\ref{eq:01}) and $|\Psi (t)\rangle $ is the wave function governed
by the the time-dependent Schr\"{o}dinger equation. By numerical
calculation, we find that for this example, when $T_{0}\geq 10/\omega
_{1}$, the system enters the adiabatic regime ($F>0.995$). 

Now in a non-adiabatic regime $T=1/\omega _{{1}}<T_{0}$ we will study the
contributions of the pulses. First, we study the effect of pulse strength for 
regular rectangular pulses.  The pulse strengths are taken to be $I=0,5\omega
_{1},8\omega _{1},10\omega _{1},20\omega _{1}$, respectively. Fig.~\ref%
{fig:1}(a) plots the fidelity as a function of $\omega _{1}t$ for
different $S$. It shows that with increasing $S$, the fidelity increases and 
$F$ approaches one for $S=0.1$ in a non-adiabatic regime.  Since the control
effect is only determined by the integral of $f(t)$, $F$ will increase with
increasing ratio $\Delta /\tau $. Fig.~\ref{fig:1}(b) shows this
property clearly.
\begin{figure}[t]
\centering
\includegraphics[width=0.8 \columnwidth]{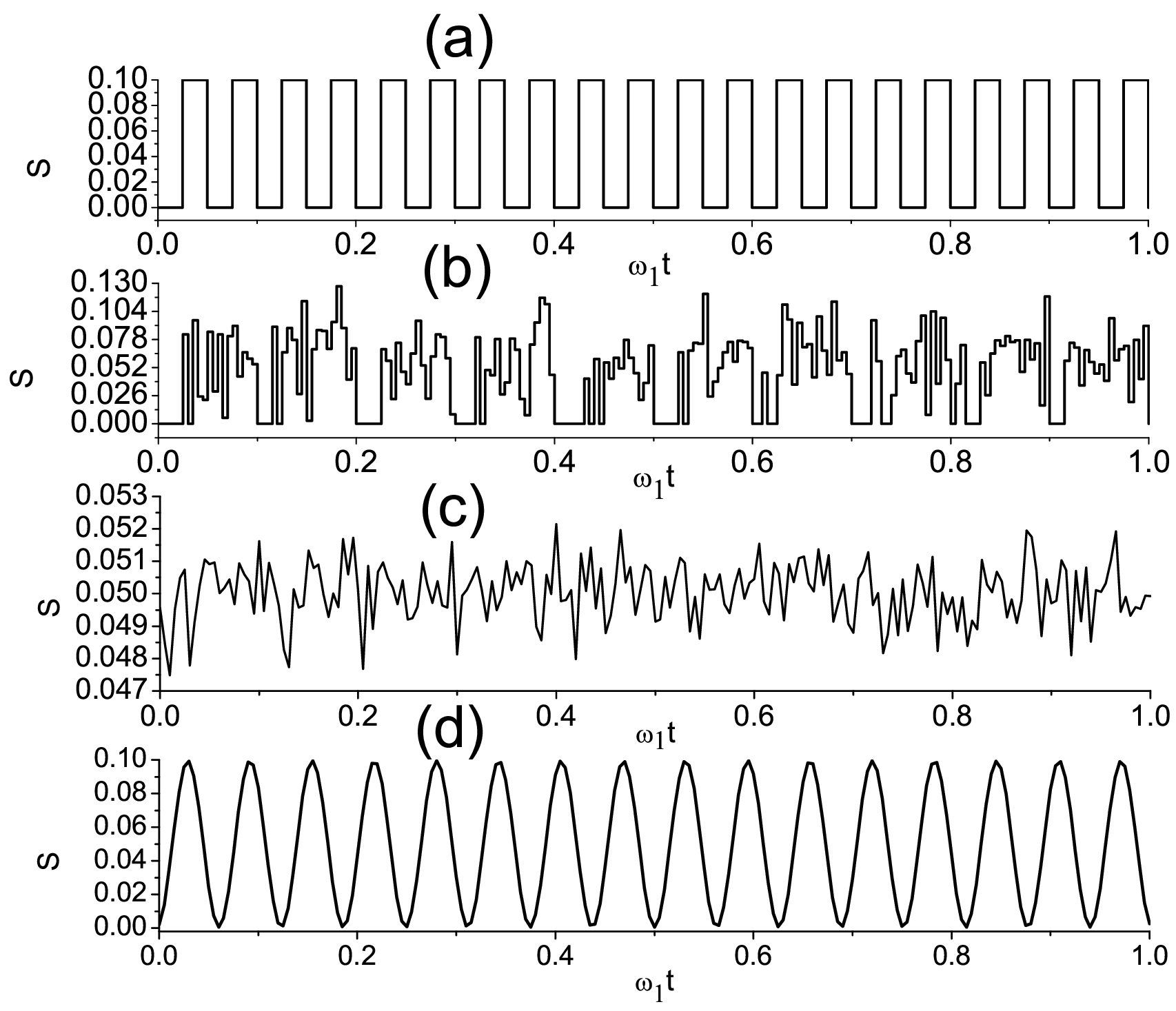}
\caption{(Color on line) Example 1: Different types of pulses, (a) regular
rectangular pulses, $\Delta/\protect\tau=5/5$, $\Delta+\protect\tau=10%
\protect\xi$, $S=0.05$; (b) random rectangular pulses,$\Delta/\protect\tau\in%
[0,1],\Delta+\protect\tau=20\protect\xi$,$S\in[0,0.1]$; (c) the average
values of a thousand times of noises with $S\in(0.047, 0.053)$; (d) a fast
sine signal $f(t)=10sin^{2}(50t)$. }
\label{fig:3}
\end{figure}

Does the pulse density affect the control result? In Fig.~\ref{fig:2}(a)-(d)
we plot the fidelity as a function of $\omega _{1}t$ for different
pulse intervals with the same ratio. Fig.~\ref{fig:2}(e) plots the
corresponding fidelity. The results show that $F$ changes slightly for
same ratio $\Delta /\tau $.

Next we consider different types of pulses.  The calculation shows that they
work as well as regular rectangular pulses \cite{JingPRL2015,WangHF}.
Suppose white noise is present \cite{JingPRA2014}, so $f(t)=\eta $rand$_{\xi }$%
(i), here rand(i) is a random number uniformly distributed in the interval $%
[0,1]$, $\eta =20\omega _{1}$. rand$_{\xi }$(i) denotes that random function
rand(i) is fixed in the time interval $\xi $, and is random for
each time interval. Fig.~\ref{fig:3}(a)-(d) plot four kinds of pulses:
regular rectangular, random rectangular, noise and a fast sine signal $%
f(t)=10sin^{2}(50t)$. Fig.~\ref{fig:3} (b) plots one possible pulse as an
example while for Fig.~\ref{fig:3}(c) we plot an average value of
the noises ($S\in(0.047,0.053)$) for different times.

\begin{figure}[t]
\centering
\includegraphics[width=0.8 \columnwidth]{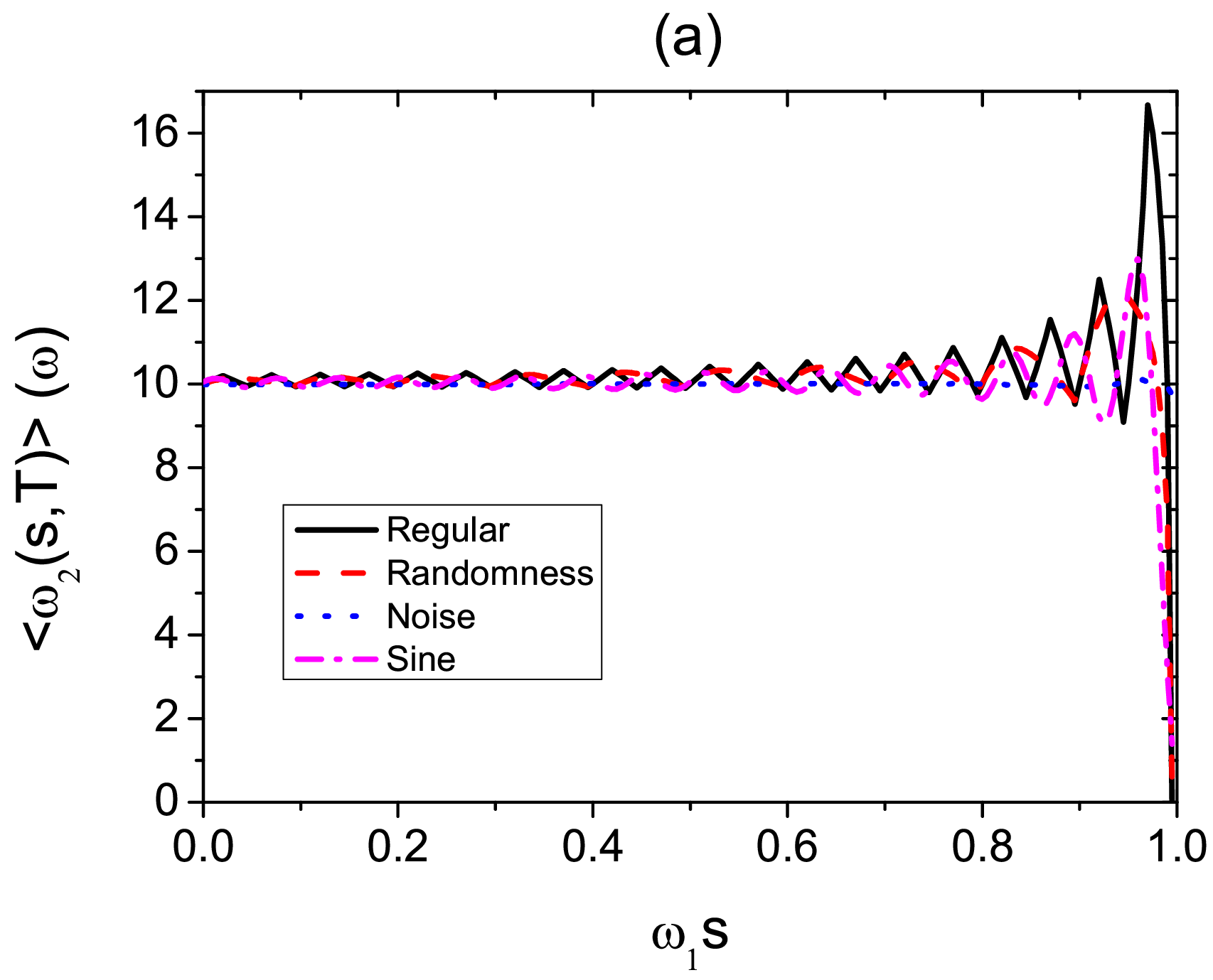} %
\includegraphics[width=0.8 \columnwidth]{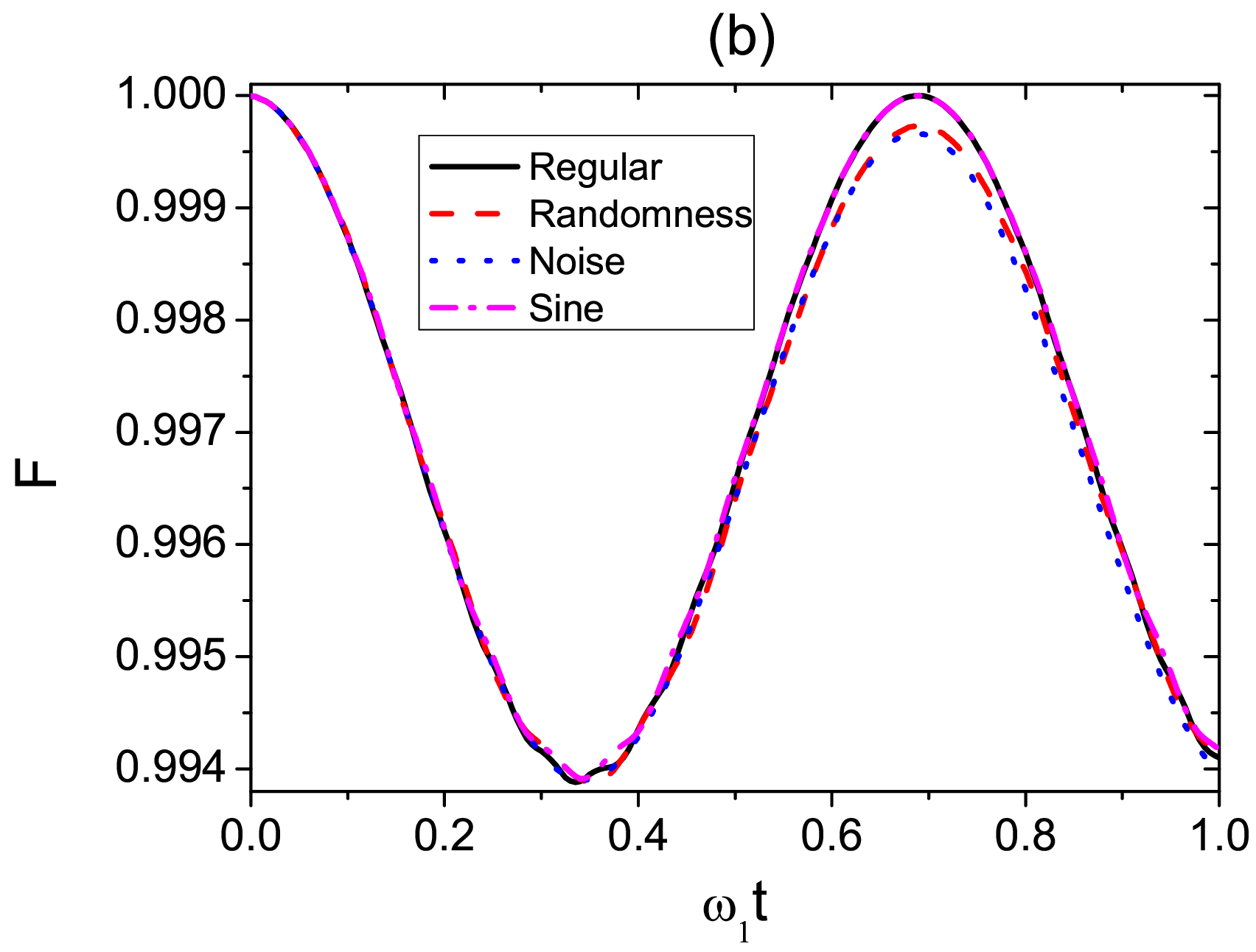}
\caption{(Color on line) Example 1: For four types of pulses plotted in Fig.~%
\protect\ref{fig:3}, (a) average control frequency versus $\protect\omega%
_{1}s$; (b) the corresponding fidelity versus $\protect\omega_{1}t$. For
randomness and noises, we calculate (thousand times) average density
matrices and obtain the corresponding average control frequency and
fidelity. }
\label{fig:31}
\end{figure}

In Fig.~\ref{fig:31}(a) and (b) we plot the corresponding average control
frequency and fidelity for four types of pulses plotted in Fig.~\ref{fig:3}.
For randomness and noise, we plot the average control frequency and
fidelity for a thousand trials.  For the four cases, the average
control frequency $\langle \omega _{2}(s,T)\rangle\approx 10\omega _{1}$,
except that near $s \rightarrow T$, there exists a fast oscillation. Fig.~%
\ref{fig:31}(b) shows that the evolution of the fidelity does not change
significantly for near equal average control frequencies. $F(1)\approx
0.994$ for the four cases. Experimentally, producing exactly regular
rectangular pulses might not be easy. Our results again relax constraints on
experimental implementation of the pulses, whether they are regular, random
and even noisy pulse sequences.

Now we turn to Example 2, a 3-spin Heisenberg XY model. The Hamilton is given
by%
\begin{equation}
H_{0}(t)=\sum\limits_{i=1}^{2}J_{i,i+1}(t)(X_{i}X_{i+1}+Y_{i}Y_{i+1})+\sum%
\limits_{i=1}^{3}h_{i}(t)Z_{i},
\end{equation}%
where $J_{i,i+1}(t)$ is the coupling between nearest-neighbor sites. Suppose
the coupling changes as $J_{i,i+1}(t)=J\sin (\Omega t)$ and the external
field changes as $h_{i}(t)=h_{i}\cos (\Omega t)$, $\Omega =\pi /(2T)$. Due
to $[\sum\nolimits_{i}Z_{i},H_{0}(t)]=0$, the magnon is conserved in the
evolution.  Therefore we only need to discuss the single-excitation
subspace where the total number of magnon is one. For simplicity, we take $%
J=\omega _{1}$ and $h_{1}=\omega _{1},h_{2}=0,h_{3}=-\omega _{1}.$ The
Hamiltonian reads%
\begin{equation}
H_{0}(t)=\omega _{1}\left(
\begin{array}{ccc}
\cos (\Omega t) & \sin (\Omega t) & 0 \\
\sin (\Omega t) & 0 & \sin (\Omega t) \\
0 & \sin (\Omega t) & -\cos (\Omega t)%
\end{array}%
\right) .
\end{equation}

The instantaneous eigenstates $| E_{n}(t)\rangle $ are $| E_{0}(t)\rangle
=\cos ^{2}(\Omega t/2)| 0\rangle -\sin (\Omega t)/\sqrt{2}| 1\rangle +\sin
^{2}(\Omega t/2)| 2\rangle ,| E_{1}(t)\rangle =\sin (\Omega t)/2| 0\rangle +%
\sqrt{2}\cos (\Omega t)/2| 1\rangle -\sin (\Omega t)/2| 2\rangle$, $|
E_{2}(t)\rangle =\sin ^{2}(\Omega t/2)| 0\rangle +\sin (\Omega t)/\sqrt{2}|
1\rangle +\cos ^{2}(\Omega t/2)| 2\rangle $ with eigenvalues $|
E_{0}(t)\rangle =-\sqrt{2}\omega _{1},| E_{1}(t)\rangle =0$ and $|
E_{2}(t)\rangle =\sqrt{2}\omega _{1}$.

The Hamiltonian in an adiabatic framework with an added LEO control now takes the form
\begin{equation}
H_{0}(t)=\left(
\begin{array}{ccc}
0 & -i\Omega /2e^{-i\sqrt{2}\omega _{1}t} & 0 \\
i\Omega /2e^{i\sqrt{2}\omega _{1}t} & 0 & -i\Omega /2e^{-i\sqrt{2}\omega
_{1}t} \\
0 & i\Omega /2e^{i\sqrt{2}\omega _{1}t} & f(t)%
\end{array}%
\right) .
\end{equation}

\begin{figure}[t]
\centering
\includegraphics[width=0.8 \columnwidth]{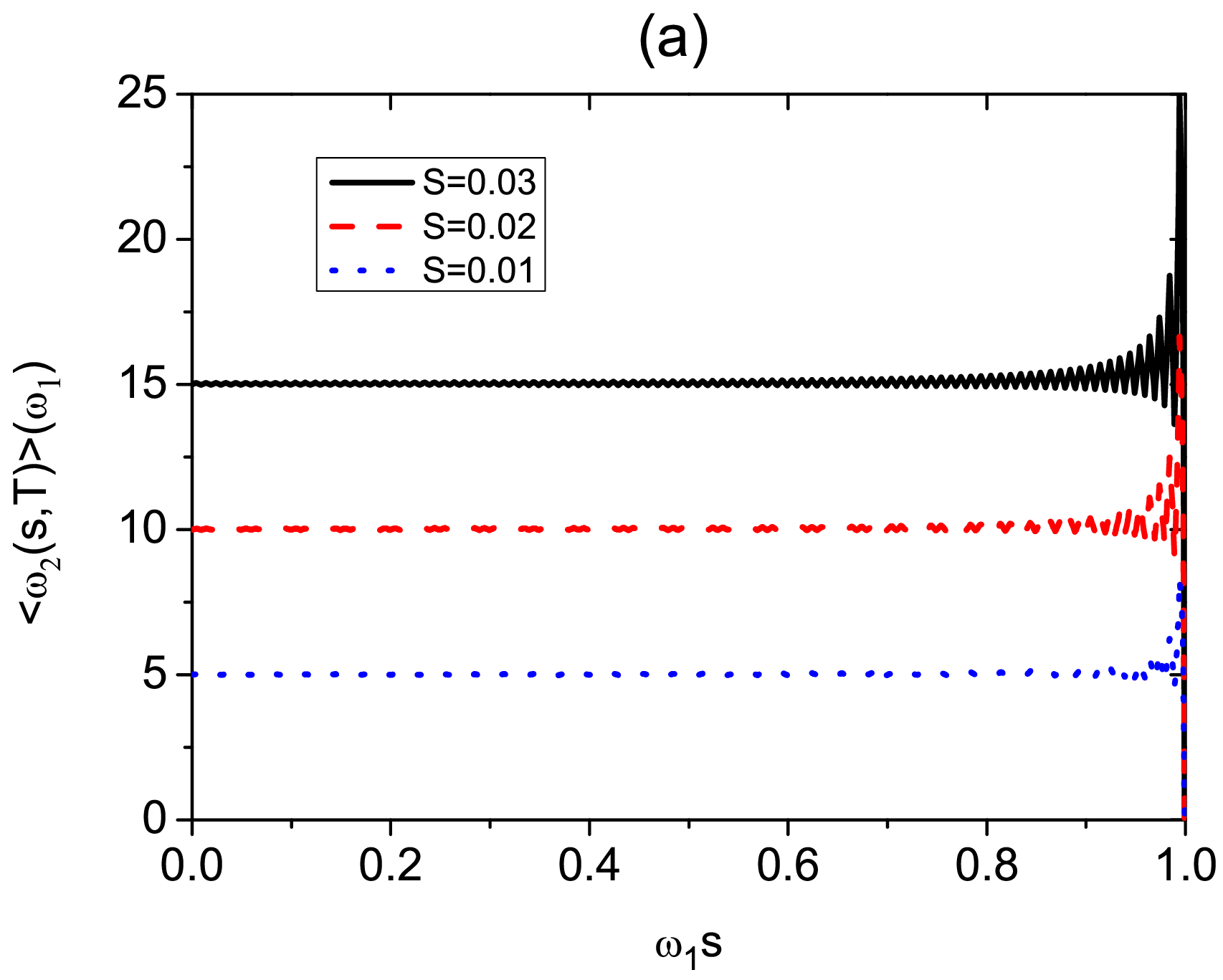} %
\includegraphics[width=0.8 \columnwidth]{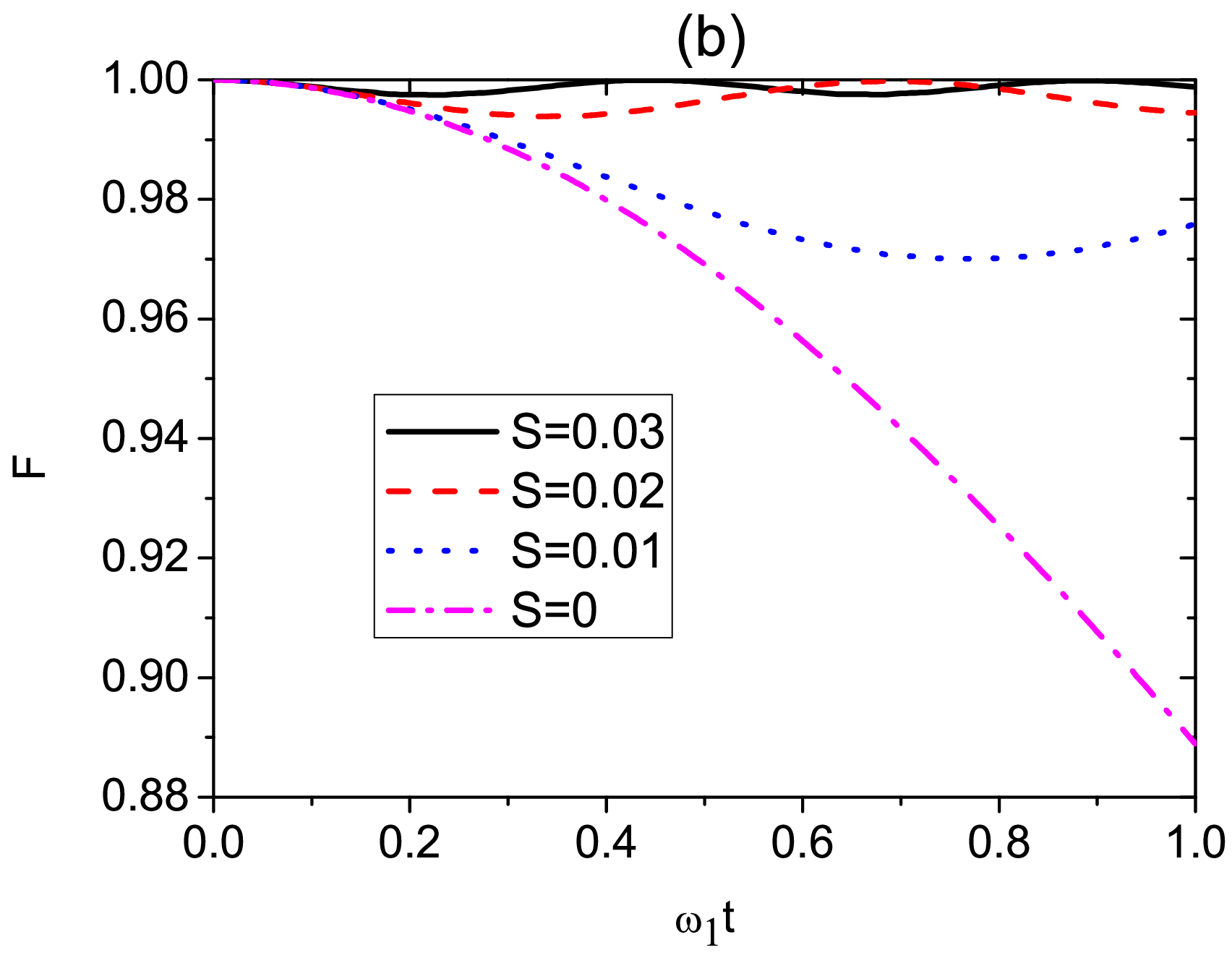}
\caption{(Color on line) Example 2: (a) The average control frequency $\langle
\protect\omega_{2} \rangle$ as function of parameter $\protect\omega_{1}s$;
(b) The corresponding fidelity versus parameter $\protect\omega_{1}t$ for
different integral $S$, $\Delta=5\protect\xi,\Delta/\protect\tau=1/1,T=1/%
\protect\omega_{1}$, the time interval we used for the calculations are taken to be 
$\protect\xi=0.001/\protect\omega_{1}$;}
\label{fig:4}
\end{figure}

The propagator can be calculated as
\begin{equation}
g^{\prime }(t,s)=-\frac{\Omega ^{2}}{4}\cos [\frac{\Omega (t-s)}{2}]\exp
\{i[\langle \omega _{2}(s,t)\rangle -\omega _{1}](t-s)\},  \label{Eq18}
\end{equation}%
where $\langle \omega _{2}(s,t)\rangle $ is the average control frequency
defined in Eq.~(\ref{Eq11}). As in the first example, when $\Omega
\rightarrow 0$, $g^{\prime }(t,s)\rightarrow 0$, the standard adiabatic
conditions are obtained. Compared with Example 1, the propagator $g^{\prime
}(t,s)$ is tuned by a cosine function $\cos [\Omega (t-s)/2]$. The
adiabaticity can also be enhanced by increasing the average control
frequency. For big $\langle \omega _{2}\rangle -\omega _{1}$, the
quickly varying factor is $\exp [i(\langle \omega _{2}\rangle -\omega
_{1})(t-s)]$ and the slowly varying factor is $\cos [\Omega (t-s)/2]p(s)$.
The quickly varying factor eliminates all the off-diagonal elements of the
propagator and effective adiabaticity is obtained.

For this example, we consider the non-adiabatic regime where $T=1/\omega _{1}$.
Fig.~\ref{fig:4} shows (a) a plot of the average control frequency $\langle \omega
_{2}(s,T)\rangle $ as a function of parameter $\omega _{1}s$ for different $%
S $. Fig.~\ref{fig:4}(b) a plot of the corresponding fidelity. In Fig.~\ref%
{fig:4}(b) the dash-dotted curve depicts the fidelity without external
pulses. It decays monotonically with time. Clearly when $\langle \omega
_{2}\rangle \approx 10$ $\omega _{1}$, $F>0.993$. When $\langle \omega
_{2}\rangle \approx 15$ $\omega _{1}$, $F>0.995$, and effective adiabaticity is
induced.

\section{LEO in experimental framework}

The analysis presented above clearly shows that the LEO in an adiabatic
frame can be used to prevent transitions.
Thus effective adiabaticity is obtained in a non-adiabatic regime. However, the
control we add is in the adiabatic frame.  What is the experimental manifestation?  
To see this, we transform to the lab frame.  For Example 1,
\begin{eqnarray}
UH_{LEO}U^{\dag } &=&f(t)[\cos ^{2}\omega t/2| 0\rangle \langle 0|  \notag \\
+\sin ^{2}\omega t/2 &|&1\rangle \langle 1| ]  \notag \\
-(\sin \omega t)/2( &|&1\rangle \langle 0| +| 0\rangle \langle 1| ).
\end{eqnarray}

For Example 2,
\begin{equation}
UH_{LEO}U^{\dag }=f(t)| \Phi (t)\rangle \langle \Phi (t)|,
\end{equation}%
where $| \Phi (t)\rangle =U| E_{0}(t)\rangle =[1-\sin ^{2}(\Omega t)/4]|
0\rangle -3\sqrt{2}\sin (2\Omega t)/8| 1\rangle +3\sin ^{2}(\Omega t)/4$ $|
2\rangle $. That is to say, if we apply the above pulse control in lab
frame, then it is equivalent to adding an LEO in an adiabatic frame.

\section{Conclusion}

Reducing the runtime for quantum information processing tasks is of
crucial importance for improving performance. We have
introduced an effective control scheme to speed up adiabatic passage by 
adding an LEO in an adiabatic framework. LEOs \cite{WuPRL2002} 
are general and 
can be applied to subspaces or subsystems \cite{Byrd} by using 
logical operations for the LEO.  Here we have shown that 
for our two examples, the $PQ$
partitioning technique can be used to derive an analytic solution for maintaining 
the system in an instantaneous eigenstate. Numerical calculations
show explicitly that the average control frequency, rather than the
details of the control function, determines the control effect \cite{JingPRL2015}. This greatly
relaxes the constraints of applying regular pulses in experiments. More importantly the control function in an experimental framework is given.  This can be applied in the field
of adiabatic quantum information processing to improve performance for 
adiabatic algorithms, 

\begin{acknowledgements}
We thank \'I\~nigo L. Egusquiza for his useful comments. This material is based upon work supported by NSFC (Grant Nos. 11475160, 61575180,11575071) and the Natural Science Foundation of Shandong Province (Nos. ZR2014AM023, ZR2014AQ026), and the Basque Government (grant IT472-10), the Spanish MICINN (No. FIS2012-36673-C03-03).
\end{acknowledgements}

\end{document}